\documentclass[12pt]{article}
\usepackage[T1]{fontenc}
\usepackage{amssymb,amsthm,amsmath,mathrsfs,bm}

\begin{document}

\title{A four component  cubic  peakon (4CH) equation}
\author{Ziemowit Popowicz}
\maketitle

\begin{center}{
Institute of Theoretical Physics, University of Wroc\l aw,

Wroc\l aw pl. M. Borna 9, 50-205 Wroc\l aw Poland, ziemek@ift.uni.wroc.pl}
\end{center}

\begin{abstract}
Two different  four component Camassa-Holm (4CH) systems with  cubic nonlinearity are proposed. 
The  Lax pair and Hamiltonian structure are defined for both (CH) systems.
The first (4CH) system include as a special case the (3CH) system considered by  Xia, Zhou and Qiao, while the second contains
the two-component generalization of Novikov system considered by Geng and Xiu. 
\end{abstract}

\section{Introduction}

\vspace{1cm}

Solitons and integrable models are very attractive objects in nonlinear sciences, originally found in experiments on shallow water propagations one and
half centuries ago. In recent years, the Camassa-Holm (CH) \cite{Holm} equation 
\begin{equation}\label{komuch}
 m_t + 2mu_x + m_xu=0,~~~~~~~ m=u-u_{xx},
\end{equation}
was derived with the aid of an asymptotic approximation to the Hamiltonian  for the Green-Naghdi equations \cite{Green} and  has attracted much attention. 
The most interesting feature of the Camassa-Holm equation is the admitance of a peaked soliton, the so called peakon solution.  
A peakon is a weak solution in some Sobolev space with a corner at its crest. 
Subsequently there are a large number of papers where various properties of these equation are established. For example, it is shown  that CH  equation 
is integrable by the inverse scattering transformation, it possesses tha Lax pair formalism, multisoliton solitons and multipeakon solutions $[ 3-13]$.

In 1998, Degasperis and Procesi \cite{Degasperis} proposed a different equation 
\begin{equation}\label{DP}
 m_t + 3u_xm + m_xu=0, ~~~~~~ m=u-u_{xx},
\end{equation}
which admits the peakon solution as well. 

The  solutions of the equation \ref{DP} are not mere abstractizations: 
the peakons replicate a feature that is characterizing for the waves of great height-waves of  of largest amplitude that are exact solutions 
of the govering equations for irrotational water waves \cite{w1,w2}. The Degaspersi-Procesi equation \ref{DP} is also considered as the 
horizontal component of the 
fluid velocity at the time in the spatial x-direction \cite{const1} with momentum density \textit{y}, but evaluated at the different level line of th fluid domains. 

Both the Comassa-Holm and Degasperis-Procesi equation are  third order equations with quadratic nonlinearity. 
Olver and Rosenau  in 1996 \cite{Rose}  proposed a general approach to construction of Camasa-Holm like equations with the peakon solution.
In particular,  the following equation appears 
\begin{equation}\label{ORQ}
m_t= -\big [m(u^2 - u_x^2) \big ]_x, ~~~~~m =u-u_{xx}
\end{equation}
and has been considered  by Fokas \cite{Fokas,Himon} and  Qiao \cite{Qiao} referred as Fokas-Olver-Rosenau-Qiao's  (FORQ) equation.

In 2008,  Novikov \cite{Novikov},  in his symmetry classification study,  proposed several differnt generalizations of the Cammasa-Holm equation to the  
peakon equations with cubic nonlinearity. One  of them 
\begin{equation}\label{Novik}
 m_t= -u^2m_x - 3u u_xm,
\end{equation}
referred to as the Novikov equation has been thoroughly considered by scientists from a different point of view. 

The Fokas-Olver-Rosenau-Qiao and Novikov equations, referred to sometimes as the cubic CH  equations, are integrable equations, 
which possess the Lax representation and   bi-Hamiltonian structures \cite{gas,hwang,Hone}. 
For example,  equation (\ref{Novik}) arises as a zero curvature equation $F_t - G_x + [F,G ] = 0$ ~\cite{Hone} where 
\begin{equation} \label{laxik}
 F= \left ( \begin{array}{ccc} 0 & \lambda m & 1 \\
          0 & 0 & \lambda m \\ 1 & 0 & 0 \end{array} \right), ~~~
G= \left ( \begin{array}{ccc} \frac{1}{3\lambda^2} -u_{x}u &  
\frac{u_{x}}{\lambda} - \lambda u^2 m & u_{x}^2 \\
 \frac{u}{\lambda} & - \frac{2}{3\lambda^2}   &  -\frac{u,x}{\lambda} - \lambda u^2m  \\
 -u^2 &\frac{u}{\lambda} & \frac{1}{3\lambda^2} + uu_{x} \end{array} \right) ,  
 \end{equation}

while  equation (\ref{ORQ}) arises   as a zero-curvature equation 
$U_t - V_x + [U,V]=0 $  where 
\begin{eqnarray}\label{Qialax} 
 U &=& -\frac{1}{2}\left ( \begin{array}{cc} 1 & -\lambda m  \\
          \lambda m & - 1 \end{array} \right), \\ \nonumber 
 V &=& \left ( \begin{array}{cc} \lambda^{-2} + \frac{1}{2}(u^2- u_{x}^2)  &  
-\lambda^{-1}(u - u_x)  - \frac{1}{2}\lambda  m(u^2 - u_x^2 \\
 \lambda^{-1}(u+u_x) + \frac{1}{2}\lambda m(u^2 - u_x^2) & -\lambda^{-2}   -\frac{1}{2}(u^2 - u_x^2)\end{array} \right).
\end{eqnarray}

The Camassa-Holm and cubic CH equations have several different generalizations to the two-component  case.  
In 2012   Xia, Qiao and Zhou \cite{Qiao1,Qiao2} generalized the Olver-Rosenau-Qiao  equation (\ref{ORQ})  to the two-component case
\begin{eqnarray} \label{qia1}
 m_t &=& \Big [m(u v - u_xv_x ) \Big ]_x - m (u v_x - u_x v) \\ \nonumber 
 n_t &=& \Big [n(u v - u_xv_x ) \Big ]_x + n (u v_x - u_x v) \\ \nonumber 
 m &=& u - u_{xx}, ~~~ n= v - v_{xx}
\end{eqnarray}
and presented its Bi-hamiltonian structure and Lax representation. 
The spectral part of the Lax representation for  equation (\ref{qia1}) is 
\begin{equation}\label{Qiaolax1}
 U = -\frac{1}{2}\left ( \begin{array}{cc} 1 & -\lambda m  \\
          \lambda n & - 1 \end{array} \right), 
\end{equation}

The Novikov equation (\ref{Novik}) has been extended to the two-component case 
\begin{equation} \label{NowyLL}
 m_{1,t} = u_1u_2m_{1,x} - 3u_2u_{1,x}m_1, ~~~~~ m_{2,t} = -u_1u_{2}m_{2,x} - 3 u_1u_{2,x}m_2 ,
\end{equation}
where $m_i=u_i-u_{i,xx}, ~ i=1,2$ by Geng and  Xue \cite{Geng}. This system has  Bi-Hamiltonian formulations and possesses
the Lax representation \cite{Li}.

Further generalizations to the multicomponent cubic Camassa-Holm equation have been considered in the literature. 
For example,  Qia \cite{QU1} generalized the Lax representation (\ref{Qiaolax1}) in which  $n,m$ are $N$ dimensional vector function. 
On the other side  Xia, Zhou and Qia \cite{QU2} generalized the Lax representation (\ref{Qiaolax1}) in which   $m=n$ is now  a  two dimensional matrix.
They have shown that the system  of equation obtained from this matrix Lax representation, referred to as 3CH system, 
is a Hamiltonian one and can be presented as
 
\begin{eqnarray}\label{Qian1}
 m_t &=& \frac{1}{2}[m(u^2 - u_x)]_x + \frac{1}{4}[m(u u_x - u_xu) - (uu_x - u_xu)m] \\ \nonumber 
u &=& \left ( \begin{array}{cc} u_{1,1} & u_{1,2}\\ u_{2,1} & - u_{1,1}  \end{array} \right ), 
~~~~  m = u - u_{xx}.
\end{eqnarray}

Moreover recently the Lax representation (\ref{laxik})  has been generalized \cite{PopLiu} to the four component case as 

\begin{eqnarray} \label{lax1}
 \Psi_x & =&  \left  ( \begin{array}{ccc} \Psi_1 \\  \Psi_2 \\ \Psi _3\end{array} \right )_x = \left(\begin{array}{ccc} 0  &  \lambda m_1  & 1  \\
&& \\\lambda m_3 & 0   & \lambda m_2 \\& & \\1 & \lambda m_4    &  0
\end{array} \right ) \left  ( \begin{array}{ccc} \Psi_1 \\  \Psi_2 \\ \Psi _3\end{array} \right ),\\ \nonumber 
\Psi_t &=& \left  ( \begin{array}{ccc} \Psi_1 \\  \Psi_2 \\ \Psi _3\end{array} \right )_t = 
\left(
    \begin{array}{ccc}
      -g_{1}f_{1} & \frac{g_{1}}{\lambda}  -\lambda m_1 \Gamma & -g_{1}g_{2} \\
      \frac{f_{1}}{\lambda} -\lambda m_3\Gamma   & -\frac{1}{\lambda^{2}}+f_{1}g_{1}+g_{2}f_{2} & \frac{g_{2}}{\lambda}
-\lambda m_2\Gamma \\
      -f_{2}f_{1} & \frac{f_{2}}{\lambda} - \lambda m_4\Gamma & -f_{2}g_{2} \\
    \end{array}
  \right)  \left ( \begin{array}{ccc} \Psi_1 \\ \Psi_2 \\ \Psi _3\end{array} \right )
\end{eqnarray}
where $\Gamma$ is an arbitrary function and $ f_1 = u_2 - u_{3,x}, ~~ f_2 = u_1 + u_{4,x}, \\
 g_1 = u_4 + u_{1,x}, ~~ g_2 = u_3 - u_{2,x},~~ m_i=u_i - u_{i,xx},~~ i =1,2,3,4 $

The integrability condition $\Psi_{x,t} = \Psi_{t,x}$ leads us 
to the system which generalizes the cubic  Comassa-Holm type equation with an arbitrary function $\Gamma$.  
The freedom in the choice of the function $\Gamma$ allows us to recover many known cubic (CH),  as well as CH type equations. 
In the special reduced cases, when $\Gamma$ is a bilinear form in $u,u_x$,    these equations contain: the three-component  system proposed by Geng and Xue \cite{Geng1} 
and two-comopnent Novikov's equation \ref{NowyLL}. In this sense almost all known $3 \otimes 3$ spectral problems for 
the CH type equations are contained for this.

In this paper, we would like to investigate the equations of motion and Hamiltonian structure obtained from the matrix Lax representation (\ref{lax1}).
By matrix representation we mean that in the Lax representation (\ref{lax1})  $m_i$ are now  $N$ dimensional matrices.
We consider the case of two dimensional matrices for which as we will show, it is possible to obtain 
from the Lax representation two different four-component cubic peakon equations. We referred these equations as 4CH equations. 

The first (4CH) system includes, as a special case, the (3CH) system considered by  Xia, Zhou and Qiao \cite{QU2}, while the second contains the two-component
generalizations of the Novikov system considered by Geng and Xue \cite{Geng}. For these  equations, we have constructed   the Hamiltonian operators.

The paper is organized as follows. In the first section, we present the general form of the matrix cubic peakon system obtained from the matrix Lax representation
 (\ref{lax1}). In the next sections, we consider the special reduced version, where we deal with the two-dimensional matrices $m_i$. In subsection $A$, 
 we consider the case $ m_3=m_2, 
m_4=m_1$, while, in  subsection $B$, the case $m_3=m_4=0$. For both cases, the Hamiltonian structure is 
constructed. The paper contains two appendices,  in which we prove that our Hamiltonian operators satisfy the Jacobi identities. The last section  contains   concluding remarks.

\section{Matrix peakon equation.} 
Let us consider the matrix version of the Lax representation equations (\ref{lax1}),  in which 
$m_i = u_1 - u_{i,xx}, i=1,2,3,4$  are $N$  dimensional matrices, $\Gamma$ is an arbitray scalar function, not a matrix.

The integrability  condition $\Psi_{x,t} = \Psi_{t,x}$ forces the following  equation of motion
\begin{eqnarray} \label{cztery}
&& m_{1,t}= -(m_1\Gamma)_x + m_4\Gamma - m_1(g_2f_2 + f_1g_1)  - g_1(g_2m_4 + f_1m_1), \nonumber\\
&& m_{2,t}= -(m_2\Gamma)_x - m_3\Gamma  +m_{2}f_{2}g_{2}+m_{3}g_{1}g_{2}+(g_2f_2+f_1g_1)m_2,\nonumber\\
&& m_{3,t}= -(m_3\Gamma)_x - m_2\Gamma  + (f_{1}g_{1}+ g_{2}f_{2})m_3 +(m_2f_2 + m_3g_1)f_1,\nonumber \\
&& m_{4,t}= -(m_4\Gamma)_x + m_1\Gamma - f_{2}(f_{1}m_1+g_{2}m_4) -m_4(f_{1}g_{1}+g_{2}f_{2}) 
\end{eqnarray}
where $f_1,f_2,g_1,g_2$ are defined in the same manner as in the scalar case but now $u_i,i=1,2,3,4$ are matrices. 

Our equations (\ref{cztery}) depend on four arbitrary $N$ dimensional matrix functions $m_i, i=1,2,3,4$, 
and, hence constitute the system of $4N$ equations. 

The Hamiltonian structure 
for our equations, if such exists, depends on these arbitrary functions as well. In the non-matrix case, we can assume a special 
reduction of the function $m_i, i=1,2,3,4$, fix the $\Gamma$ function,  and then 
find the Hamiltonian structure.
In the next section, we will show that for the two-dimensional matrices it is also possible to assume special reduction of matrices $m_i$,  fix  $\Gamma$ function,  and finally  find the Hamiltonian structure.

\section{4CH equations.}
In this section,  we consider the case where  $m_i$ are two dimensional matrices only. In this situation,  we have 16 arbitrary functions which
parametrize the matrices $m_i$,  and 
the equations (\ref{cztery}) constitute 16 equations. It appear that, when  we restrict the considerations to the four arbitrary functions,
it is possible to fix the $\Gamma$ function 
and find the Hamiltonian formulations  for the system of equations (\ref{cztery}).

\subsection{First 4CH equation  $m_3=m_2,m_4=m_1$.}

For this case our equation (\ref{cztery}) reduces to 
\begin{eqnarray} 
 m_{1,t} &=& -(m_1\Gamma)_x + m_1\Gamma + \\ \nonumber 
&&  2 m_1 (u_{2,x}- u_2)(u_{1,x} +u_{1}) +  2 (u_{1,x} + u_{1} ) (u_{2,x}  - u_{2}) m_1 \\ \nonumber
 m_{2,t} &=& -(m_2\Gamma )_x - m_2\Gamma  + \\ \nonumber
 && 2 m_2 (u_{1,x}+ u_1)(u_{2} - u_{2,x}) + 
2 (u_{2} - u_{2,x} ) (u_{1,x}  + u_{1}) m_2 
\end{eqnarray}
and constitute the system of 8 equations in which $\Gamma$ is an arbitrary function. 

Further, we assume that $m_2=m_1^{\star}, u_2=u_1^{\star}$ where $\star$ denotes the hermitean conjugation,
and we parametrize our matrices $u_i$ and $m_i,~~i=1,2,$ as
\begin{eqnarray} \label{para}
 u_{1} &=&  u_{1,0} + i u_{1,1} \sigma_1 + i u_{1,2}\sigma_2 + i u_{1,3}\sigma_3,\\ \nonumber 
 u_{2} &=& u_{1,0} - i u_{1,1}\sigma_1 - i u_{1,2}\sigma_2 - i u_{1,3}\sigma_3 \\ \nonumber   
  m_k &=& u_k - u_{k,x}, ~~~~ m_{k,j}=u_{k,j} - u_{k,j,xx}, ~~~ k=1,2,~~j=0,1,2,3   
\end{eqnarray}
where $\sigma_1$ are Pauli matrices. 

From the assumption $m_{2,t} = m_{1,t}^{\star}$ it follows that  $\Gamma = 4\sum_{k=0}^{3}(u_{1,k}^2 - u_{1,k,x}^2)$. 

Now, our equations (\ref{cztery}) reduce to the system of four equations
\begin{equation} \label{row} 
m_{1,i,t} = -(\Gamma m_{1,i})_x + \sum_{k=0}^{3} (u_{1,k,x}u_{1,i} - u_{1,k} u_{1,i,x})m_{1,k}
\end{equation}
It is our first four component cubic 4CH system.

Apparently, this system  reduces to the matrix peakon equation (\ref{Qian1}), when 
\begin{eqnarray}\label{piczki}
 &&u_{1,0}= 0,~~~ u_{1,1}= v_1,~~~u_{1,2}=\frac{v_2+v_3}{2},~~~ u_{1,3}=-i\frac{v_2 - v_3}{2}, \\ \nonumber 
&& \hspace{2cm} m_{1,i}=v_i - v_{i,xx},~~~~ i=1,2,3
\end{eqnarray}
It can be  further reduced to the Camassa-Holm equation (\ref{komuch}), when $v_3=0,v_1=0$, or to the cubic CH equation, when $v_3=v_4=0$. 

In order to obtain  the Hamiltonian formulation of the first  cubic peakon  4CH system,  let us  introduce 
new a parametrization of $u_{1,j}, m_{1,j}$  as
\begin{eqnarray} 
u_{1,0} &=& \frac{1}{2}(v_1 + v_2) ,~~~~~  u_{1,1} = -\frac{i}{2}(v_1 - v_2) \\ \nonumber 
 u_{1,2} &=& \frac{1}{2}(v_3 + v_4) ,~~~~~  u_{1,3} = -\frac{i}{2}(v_3 - v_4)  \\ \nonumber 
n_j=v_j - v_{j,xx}, ~~~~\Gamma &=& 4(v_1v_2 + v_3v_4 - v_{1,x}v_{2,x} - v_{3,x}v_{4,x}),  \nonumber
\end{eqnarray}
where $j=1,2,3,4$ 
and our equation (\ref{row}) transforms to  
\begin{eqnarray}
n_{i,t} &=& -(n_i \Gamma)_x + 4v_i \big (v_{1,x}n_2 + v_{2,x}n_1 + v_{3,x}n_4 + v_{4,x}n_3 \big ) - \\ \nonumber 
&& \hspace{2cm} 4v_{i,x}\big ( v_1n_2 + v_2n_1 + v_3n_4 + v_4n_3 \big ).
\end{eqnarray}

These equations  could be formulated as  the  Hamiltonian system,
\begin{eqnarray}
H &=& \frac{1}{2} \int ~dx ~ (v_1 n_2 + v_3 n_4  ), \hspace{2cm}   {\cal L} = {\cal L}_1 -{\cal L}_2   \\ \nonumber  
 {\cal L}_{1,i,j} &=& - \partial n_i \partial^{-1} n_j \partial  +  \delta_{i.j} n_i
 \partial^{-1} n_j\\ \nonumber 
&& ~~~\\ \nonumber 
{\cal L}_2 &=& \left ( \begin{array}{cccc}  0 &
  \begin{array}{ccc}     n_4 \partial^{-1} n_3 +\\
          n_3\partial^{-1}n_4 + \\ n_1\partial^{-1}n_2 \end{array} &
 -n_3 \partial^{-1} n_1 &  -n_4\partial^{-1}n_1 \\ 
 \begin{array}{ccc}  n_4\partial^{-1}n_3 +  \\
 n_3\partial^{-1}n_4 + \\ n_2\partial^{-1}n_1 \end{array} 
& 0 & -n_3\partial^{-1}n_2 &
-n_4\partial^{-1} n_2  \\-n_1\partial^{-1}n_3 & -n_2\partial^{-1}n_3 & 0 & \begin{array}{ccc} n_2 \partial^{-1} n_1+  \\
n_1\partial^{-1}n_2 +\\ n_3\partial^{-1}n_4\end{array} 
 \\-n_1\partial^{-1}n_4  & -n_2\partial^{-1}n_4 &  \ \begin{array}{ccc} n_2 \partial^{-1} n_1 +\\
n_1\partial^{-1}n_2 + \\n_4\partial^{-1}n_3 \\ \end{array}       & 0 \end{array} \right ) 
\\ \nonumber 
\end{eqnarray}
\begin{equation} \label{dwa}
\left ( \begin{array}{cccc} n_1 \\ n_2 \\ n_3 \\ n_4 \end{array} \right )_t = 4 {\cal L} 
\left ( \begin{array}{cccc} H_{n_1} \\ H_{n_2}\\H_{n_3}\\H_{n_4} \end{array} \right ),
\end{equation}
where $ H_{n_i} = \frac{\delta H}{\delta n_i}$, and 
\

The proof that the  Jacobi identity for the ${\cal L}$ operator holds is delegated  to  appendix A.

\subsection{ Second 4CH equation  $m_3=m_4=0$.} 

Similarly to the first  cubic peakon  4CH system we assume that $m_2=m_1^{\star}$ and parametrize the matrices 
$u_i$ and $m_i,~i=1,2$ as
\begin{eqnarray}
u_1 &=& r_1 + ir_2\sigma_1 + ir_3\sigma_2 + ir_4\sigma_3,~~~~~
u_2 = r_1 - ir_2\sigma_1 - ir_3\sigma_2 - ir_4\sigma_3 \\ \nonumber 
m_1 &=& p_1 + ip_2\sigma_1 + ip_3\sigma_2 + ip_4\sigma_3,~~~~~
m_2 = p_1 - ip_2\sigma_1 - ip_3\sigma_2 - ip_4\sigma_3 \\ \nonumber 
m_i &=& u_i - u_{i,xx},~~ i=1,2, \hspace{1.9cm} p_j=r_j- r_{j,xx},~~~~ j=1,2,3,4 \\ \nonumber 
\end{eqnarray}
The assumption $m_{3,t}=m_{4,t}=0$  forces that  $\Gamma = r_1^2 + r_2^2 + r_3^2 +r_4^2$. 

Because if this ,  the equations  (\ref{cztery}) become 
\begin{eqnarray} \label{matNovi} 
m_{1,t} &=&  m_1(u_{2,x}u_1 - u_2u_{1,x}) -   u_{1,x}u_2 m_1  - (u_1u_2m_1)_x, \\ \nonumber
m_{2,t} &=& (u_{2}u_{1,x} - u_{2,x}u_{1})m_2   - m_2 u_1 u_{2,x}  -(m_2u_1u_2)_x   
\end{eqnarray}

and it can be rewritten as
\begin{equation}\label{nowyk}
 p_{i,t} = -(p_i\Gamma)_x - \frac{1}{2}p_i\Gamma_x - \sum_{j,k,s=1}^{4}\epsilon_{i,j,k,s}p_kr_jr_{s,x} + 
3\sum_{k=1}^{4} p_k(r_ir_{k,x} - r_{i,x}r_k),
\end{equation}
where $\epsilon_{i,j,k,l}$ is the antisymmetric tensor such that $\epsilon_{1,2,3,4}=1$. 

This  is our second four component cubic 4CH  system. 

The system of equation (\ref{nowyk}) allows for further reduction. For example, in the  case when $p_3=p_4=r_3=r_4=0$ and assuming that 
$p_1 = a_1+ a_2,~~p_2=i (a_1-  a_2)$~
the system (\ref{nowyk}) reduces to  the two-component Novikov equation considered by Geng, Xiu, Li and Liu
\cite{Geng,Li}.

The system of equations (\ref{nowyk}) is a hamiltonian system,
\begin{eqnarray} 
 \left ( \begin{array}{cccc} p_1 \\ p_2 \\ p_3 \\p_4  \end{array} \right )_t = {\cal L}
 \left ( \begin{array}{ccc}  H_{p_1} \\ H_{p_2}\\ H_{p_3} \\ H_{p_4}  \end{array} \right )
\end{eqnarray}
where 
\begin{eqnarray}\label{hamik}
{\cal L}_{j,k} &=& (3p_j\partial + 2p_{j,x})(\partial_{xxx} - 4\partial)^{-1}(3p_k\partial + p_{k,x})  \\ \nonumber 
&& + \sum_{s,n=1}^{4} \epsilon_{j,k,s,n}p_s\partial^{-1}p_n - 3\delta_{j,k}\sum_{s=1}^{4} p_s\partial^{-1}p_s + 
3p_k\partial^{-1}p_j \\ \nonumber 
H &=& \frac{1}{2} \int ~ dx ~ (p_1r_1 + p_2r_2 + p_3r_3 + p_4r_4).
\end{eqnarray}

\noindent The proof that the operator satisfies the Jacobi identity is described in  appendix B.

\section{Conclusion} 
In this paper we studied   two different  four component Camassa-Holm (4CH) systems with  cubic nonlinearity.
The Lax pairs and Hamiltonian structure have been proposed for these two different systems. 
The first (4CH) system include as a special case the (3CH) system considered by  Xia, Zhou and Qiao while the second contains the two-component 
generalizations of Novikov system considered by Geng and Xiu. 
Our Lax pair is a matrix generalization of the Lax pair for the four component 
Camassa-Holm type hierarchy considered in \cite{PopLiu}.

Our matrix Lax representation produces a 
huge number of different cubic CH type equations and it will be interesting to investigate  these further,
especially to study the existence of infinitely many conservation laws.

\section{Appendix A}

We use  traditional manner to verify the Jacobi identity \cite{Blasz}.
In  order to prove  that the operator $\cal L$, defined in (\ref{dwa}),  satisfies the Jacobi identity, we utilize the standard form of the Jacobi identity
\begin{equation} \label{Jacob}
 Jacobi = \int~dx A {\cal L}^{\star}_{{\cal{L}}(B)} C + cyclic(A,B,C) = 0,
\end{equation}
where $A,B$ and $C$ are the test vector  functions for example  $A=(a_1,a_2,a_3.a_4)$ while $\star$ denotes the Gato derivative along the vector ${\cal{L}}(B)$.
We check this identity utilizing the computer algebra Reduce and package Susy2 \cite{Sus}. 
We will, briefly  explain our procedures used  during the verification of the Jacobi identity (\ref{Jacob}).

In the first stage, we remove the derivatives from the test functions $a_{i,x},b_{i,x}$ and $c_{i,x}$ in the Jacobi identity, using the rule 
\begin{equation} \label{tiki}
f_x = \overset{\leftarrow}{\partial} f - f \vec \partial 
\end{equation}
where $f$ is an arbitrary function. 

Because of this,  the Jacobi identity   can be split into three  segments. The first and second segments  contain  terms in which the  integral 
operator $\partial^{-1}$ appears twice and once respectively. The last segment does not contain any integral operators. We consider each segment  separately. 

The first segment is a combination of the following expressions:
\begin{equation} \nonumber   \int ~dx~ n_{c}\partial^{-1} n_{a} \partial^{-1}n_{b}  + 
 \int ~dx~ \tilde  n_{a} \partial^{-1} \tilde n_{c} \partial^{-1} \tilde n_{b} + cyclic(a,b,c)
\end{equation}
where $n_{a}$ denotes $n_ja_i$ or $n_{j,x}a_i$ or $n_{j,xx}a_i,~~~ i,j = 1,2,3,4$ and similarly for $n_{b},n_{c}, $ 
$\tilde n_a,\tilde n_b,\tilde n_c$.

Here $\partial^{-1}$ is an integral operator, and, therefore, each ingredient could be rewritten as, 
for example,  
\begin{equation} \label{trud} 
  \int ~dx~ n_{c}\partial^{-1} n_{a} \partial^{-1}n_{b} = 
 - \int ~dx~ n_{a} (\partial^{-1} n_{c})(\partial^{-1} n_{b}),
\end{equation} 

Now, we replace  $n_{a}$ in the last formula by 
\begin{equation}\nonumber
 n_{a} =  \overset{\leftarrow}{\partial} (\partial^{-1} n_{a}) - (\partial^{-1}n_{a})\vec \partial .
\end{equation}
Hence, the expression (\ref{trud}) transforms to 
\begin{equation}\nonumber 
  \int ~dx~ n_{c}\partial^{-1} n_{a} \partial^{-1}n_{b} = \int ~dx,  ~n_c(\partial^{-1}n_a)(\partial^{-1}n_b) + \int ~dx ~n_b (\partial^{-1} n_a)(\partial^{-1}n_c)
\end{equation}

Now repeating this procedure for $n_a$ and $\tilde n_a$ in  the first  segment, it appears that this segment   reduces to zero.

The second segment  is constructed from the combinations  of  the following terms:
\begin{equation}\nonumber
\int ~dx~\Lambda_{a} \Lambda_{c} \partial^{-1} \Lambda_b + 
\int ~dx~\tilde  \Lambda_{b}\partial^{-1} \tilde \varLambda_{a}\tilde \varLambda_{c} + cyclic(a,b,c).
\end{equation}
where $\Lambda_a$ takes values in  $ \{ n_{j}a_{i},~n_{j,x}a_{i},~n_{j,xx}a_{i},~n_{j}a_{i,x},~ n_{j}a_{i,xx}\},i,j=1,2,3,4  $. 
In a similar manner the $\Lambda_b,\Lambda_c,\tilde \Lambda_a,\tilde \Lambda_b, \tilde \Lambda_c$ are defined.
These terms are rewriten as
\begin{equation}\nonumber
\int ~dx~\Lambda_{a} \Lambda_{c} (\partial^{-1} \Lambda_b) - 
\int ~dx~\tilde \varLambda_{a}\tilde \varLambda_{c} (\partial^{-1} \tilde \Lambda_{b})  + cyclic(a,b,c).
\end{equation}

Next, using  rule (\ref{tiki}),  we replace  once more the  derivatives in $a_{k,x}$ and $b_{k,x}$   
in the second segment.  
After this replacement, it appears that  the second segment contains   no   term with an  the integral operator. 
Therefore,  we add this  segment to the  third  segment.

Now,  it is easy to check that this last segment  vanishes. Indeed, it is enough to use the rule (\ref{tiki})  in order to remove the derivatives from 
$a_{k,x}$ in the last segment. 

This finishes the proof.

\section{Appendix B}

We use  the idea of the decompression of the Hamiltonian operator \cite{Popo} in order to prove that the operator 
${\cal L}$ defined in (\ref{hamik}) satisfies the  Jacobi identity. To this  end, let us consider a new operator $\Lambda$
\begin{equation} \nonumber 
 \Lambda=\left (\begin{array}{ccccc} 
-(\partial^3 - 4v \partial - 2v_x), & 3p_1\partial + p_{1,x}, & 3p_2\partial + p_{2,x}, & 
3p_3\partial + p_{3,x}, & 3p_4\partial + p_{4,x} \\
3p_1\partial + 2p_{1,x}, & & & & \\
3p_2\partial + 2p_{2,x}, & & \varTheta & &  \\
3p_3\partial + 2p_{3,x}, & & & & \\
3p_4\partial + 2p_{4,x}, & & & & \end{array} \right )
\end{equation}
where $\varTheta$ is the matrix with the entries
\begin{equation} \nonumber
 \varTheta_{j,k} =\sum_{s,n=1}^{4} \epsilon_{j,k,s,n}p_s\partial^{-1}p_n - 3\delta_{j,k}\sum_{s=1}^{4} p_s\partial^{-1}p_s + 3p_k\partial^{-1}p_j 
\end{equation}

As we see,  the Hamiltonian operator $\cal L$,  defined in (\ref{hamik}),  is a Dirac reduced version of the $\Lambda$ operator when $v=1$. 
From this,  it is  enough to prove that the  operator $\Lambda$ satisfies the Jacobi identity. We verified the Jacobi identity using 
the same procedure as in  appendix A.

\end{document}